\documentclass[12pt]{iopart}
\usepackage{iopams,psfrag,graphicx}

\newcommand{\NN}{\mathbb{N}}

\begin{document}
\title{A model for bidirectional traffic of cytoskeletal motors}
\author{Maximilian Ebbinghaus and Ludger Santen}
\address{Fachrichtung Theoretische Physik, Universit\"at des Saarlandes, 66041 Saarbr\"ucken, Germany.}
\ead{\mailto{ebbinghaus@lusi.uni-sb.de}, \mailto{santen@lusi.uni-sb.de}}

\begin{abstract}
We introduce a stochastic lattice gas model including two particle species and two parallel lanes, one of which comprises exclusion interaction and directed motion while the other one shows no exclusion interaction and unbiased diffusion, thus mimicking a micotubule filament and the surrounding solution. At a high binding affinity to the filament, jam-like situations dominate the system's behaviour. We approximated the fundamental process of position exchange of two particles. In the case of a many-particle system, we were able to identify one regime in which the system is rather homogenous with only small accumulations of particles and another regime in which a significant fraction of all particles accumulates in the same cluster. Numerical data indicates that this cluster formation will occur at all densities for large system sizes. Coupling of several filaments leads to an increased cluster formation compared to the uncoupled system, suggesting that efficient bidirectional transport on one-dimensional filaments relies on long-ranged interactions and track formation.
\end{abstract}


\maketitle

\section{Introduction}
\label{s:intro}
In the past, several models for directed stochastic transport have been treated intensively, relying mostly on some variations of the asymmetric simple exclusion process (ASEP)~\cite{Derrida98a}. Amongst others these models used to examine several biological transport processes, such as biopolymerization, protein synthesis or motion of motor proteins along the cytoskeleton.

The metabolic needs of eucaryotic cells are met by the use of an efficient active transport system that acts on the microscopic length scales of the cells~\cite{Alberts02}. This intracellular transport consequently assures the survival of the cells; defects of this transport system happen to correlate with some diseases (e.g., Alzheimer's disease~\cite{Stokin05}).

The understanding of the basic properties and the interplay between cytoskeletal filaments and the motor proteins that drive the active transport is thus of high importance and a much discussed subject of research~\cite{Vilfan01,Mallik04,Singh05,Ross06,Seitz06,Caviston06,Beeg08}. The motor proteins transport intracellular cargo such as organelles or vesicles by performing stochastic motion along the filaments of the cytoskeleton~\cite{Schliwa03}. These filaments are polarized and motor proteins effectively move in only one direction along the filament by taking load-dependent steps of a multiple of the length of a filament subunit~\cite{Mallik04}. Another important feature is the processivity of proteins like kinesin and dyneins, which means that they perform several hundreds of steps along the microtubule (MT) filament before desorbing from the MT to the surrounding cytoplasm. An early variant of the ASEP which considers the finite processivity of molecular motors in an ASEP-like model has been suggested by Lipowsky et al.~\cite{Lipowsky01}. The resulting finite path length has also already been incorporated in models with Langmuir kinetics (e.g.~\cite{Parmeggiani03,Evans03}). The motor dynamics is stochastic, i.e., the motion along the filament as well as the detachment and attachment from and to the filament are random in nature. This and its elongated geometry render it suitable for an axon to be modeled by a one-dimensional stochastic lattice gas. In the past, the ASEP~\cite{Derrida98a} has been modified in different ways to include features of the biological situation (e.g., multiple filaments, local non-conservation of particles)~\cite{Parmeggiani03,Evans03,Klumpp03,Klumpp05,Wang07,Reichenbach07,Chowdhury08,Grzeschik08}, which have also led to the prediction of experimentally observable effects~\cite{Nishinari05}. In contrast, the number of models including bidirectional transport, which therefore take the opposed direction of kinesin's and dynein's motion along MTs into account, is rather limited~(e.g., \cite{Evans95,Klumpp04,Arndt98,Pronina07}). In general, these models show a tendency towards spontaneous symmetry breaking.

The main features of the model treated in this publication are the existence of two particle species that exclude each other from a one-dimensional filament with discrete binding sites. The particles can desorb from the filament and perform diffusive motion in a surrounding cytoplasm similar to~\cite{Lipowsky01}. Instead of modeling the diffusive environment explicitly as in~\cite{Lipowsky01}, we introduce a second lane where particles move diffusively. The two particle species move in opposite directions on the filament. A model of this kind has been introduced by~\cite{Arndt98,Korniss99} and discussed in the context of different applications, e.g. ant trails~\cite{Kunwar06}. Both models consider the exchange of particles on a given track. By contrast, in our approach particle exchanges are only possible via a sequence of desorption, diffusion and adsorption moves. Very recently, a similar model with only one particle species has been treated in~\cite{Tailleur08} including relative motion of the two lanes.

Our main interest will lie in the transport properties of a model with bidirectional motion. It turns out that the transport capacity of the system is esentially determined by the outflow from the largest cluster of the system. The formation of clusters is a bulk process. Therefore, we consider a system with periodic boundary conditions. Also, the choice of introducing a second lane instead of a grand-canonical reservoir coupled by Langmuir kinetics (as in~\cite{Parmeggiani03}) allows particles to remember the location on the filament from where they detached and thus introduces memory into the system.

The paper is organized as follows. In section~\ref{s:moddef}, a model in the spirit of~\cite{Klumpp04} is defined. Since blocking situations will limit the current notably, we first consider a reduced system in section~\ref{s:2part} with only one particle of each species and thereby derive an approximate expression for the current at low desorption rates. In section~\ref{s:manypart}, a system with many particles is treated. Our analysis combines analytical and numerical computation. The analytical results are based on mean field and phenomenological approaches, valid for particular sets of parameters. By coupling two systems in section~\ref{s:coupledsys}, we take a step towards the biological situation and find that the transport capacity of the model does not significantly increase. Finally, in section~\ref{s:discussion}, we summarize and discuss briefly the physiological relevance of the model.

\section{Model definition}
\label{s:moddef}

\begin{figure}
\begin{center} 
\psfrag{D}{$D$}
\psfrag{wd}{$\omega_d$}
\psfrag{wa}{$\omega_a$}
\psfrag{p}{$p$}
\includegraphics*[scale=0.8]{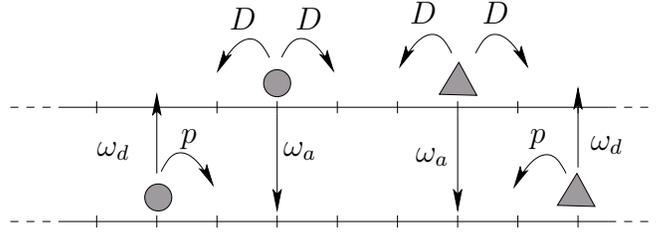}
\caption{Representation of the considered model. Grey circles and triangles refer to kinesin and dynein motors respectively. The arrows indicate the allowed moves with the corresponding rates. We impose periodic boundary conditions for both lanes, and hard-core interaction on the filament.}
\label{f:moddef}
\end{center}
\end{figure}

Similarly to \cite{Grzeschik08,Tailleur08}, we consider a two-lane lattice gas model of $L$ discrete sites with periodic boundary conditions. A schematic sketch can be found in figure~\ref{f:moddef}. The lower lane represents a microtubule filament of an axon while the upper lane symbolizes the surrounding cytoplasm (in the following called ``diffusive lane''). Molecular motors of the two most prominent MT motor protein families, kinesins (K) and dyneins (Dy), can occupy both lanes and will be referred to as ``particles''. Since a single MT protofilament offers only one binding site per tubulin subunit of length $8~nm$, we have hard-core interaction on the filament. Effectively, the occupation number is $b_i^\pm=0$ or $1$ with $i$ referring to the lattice site, $b$ to the \emph{bound} state (= lower lane) and the plus resp. minus sign to the particle type according to their preferential moving direction. The average concentration of the unbound molecular motors is expected to be low. Therefore, exclusion effects in the cytoplasm will not be considered. Consequently, there is no need to impose any restriction on the occupation number in the diffusive lane and interactions in this \emph{unbound} state will be neglected in this model, so that we have $u_i^\pm \in \NN$.

The dynamics are chosen to include the major features of intracellular transport. In the bound state, i.e., on the filament, each particle type on its own would perform a totally asymmetric exclusion process with forward hopping rate $p$. As the second species is moving in the opposite direction, encounters of particles of different species will often happen in the bound state. To relieve these blocking situations, inter-lane moves are permitted with the rates $\omega_d$ and $\omega_a$ where the indices refer to the biologically underlying desorption and adsorption processes when changing from the filament to the cytoplasm and back. Finally, in the unbound state, particles diffuse freely with the rate $D$ and, because of the absence of any interaction, perform a one-dimensional random walk until they reattach to the filament. The absence of creation or annihilation of particles and the periodic boundary conditions cause global mass conservation. The possible moves are summarized in table~\ref{t:moves}.

For the sake of simplicity, the hopping rates are chosen to be independent of the particle species, although kinesin and dynein proteins possess different dynamic properties. Choosing asymmetric hopping rates does not qualitatively affect the presented results.

Assuming the transitions to be very quick compared to the waiting times (= inversed transition rates), a random sequential update of the particles neglecting transitions with probability of the order $\rmd t^2$ is well suited for a continuous time simulation. The MC results have been obtained by running the simulation over at least $10^6$ sweeps after arriving the stationary state.

\begin{table*}
\caption{Table of possible moves ($n,m\in\NN,\ n\geq 1$) with corresponding rates in the presented model.}
\label{t:moves}
\begin{indented}
\item[]\begin{tabular}{@{}lll}
\br
Move & Rate & Biological interpretation \\
\mr
$\{b_i^+=1,b_{i+1}^+=0,b_{i+1}^-=0 \}$ & $p$ & Directed motion of kinesin along MT \\
\quad$\to \{b_i^+=0,b_{i+1}^+=1,b_{i+1}^-=0 \}$\\
\mr
$\{b_i^-=1,b_{i-1}^+=0,b_{i-1}^-=0 \}$& $p$ & Directed motion of dynein along MT \\
\quad$\to \{b_i^-=0,b_{i-1}^+=0,b_{i-1}^-=1 \}$\\
\mr
$\{u_i^\pm=n,u_{i\pm1}^\pm=m \}$& $D$ & Diffusion in cytoplasm \\
\quad$\to \{u_i^\pm=n-1,u_{i\pm1}^\pm=m+1 \}$\\
\mr
$\{b_i^\pm=1,u_i^\pm=m \}$&$\omega_d$ & Detachment from MT \\
\quad$\to \{b_i^\pm=0,u_i^\pm=m+1  \}$\\
\mr
$\{b_i^+=0,b_i^-=0,u_i^\pm=n \}$& $\omega_a$ & Attachment to MT \\
\quad$\to \{b_i^\pm=1, b_i^\mp=0,u_i^\pm=n-1  \}$ &\\
\br
\end{tabular}
\end{indented}
\end{table*}

\section{Mutual blocking: Two-particle system}
\label{s:2part}

As mentioned in the model definition, the system's behaviour will be dominated by configurations in which at least two particles of different species occupy neigbouring sites on the filament. These particles contribute further to the current along the filament (which is the physical quantity of main interest in this investigation) only if they switch sites. This exchange process is of higher importance if the detachment rate $\omega_d$ is low compared to the other transition rates. In this case, the waiting time $\omega_d^{-1}$ is the dominating time scale in the system. In order to analyse the mutual blocking of the particles, we study the elementary exchange process of two particles in detail.

\begin{figure}
\begin{center}
\includegraphics*[scale=0.8]{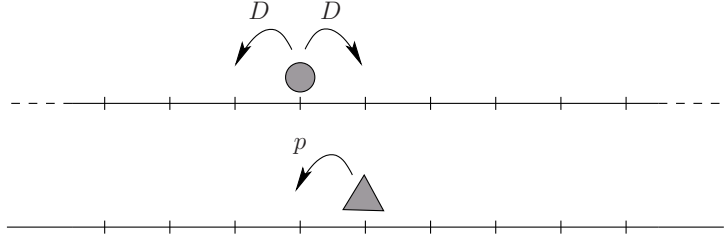}
\caption{Configuration after an initial detaching move described in section~\ref{s:2part}.}
\label{f:ptrap}
\end{center}
\end{figure}

We consider a system with only one particle of each species. Inter-lane changes happen preferentially in direction of the filament ($\omega_d/\omega_a\ll1$), so that we expect the particles to almost always be on the filament. The system periodically changes between a regime in which the particles can move freely on the filament and a regime in which the two particles try to switch sites on the filament and have no net displacement. In the free-moving state, a particle performs on average $L/2$ consecutive steps before encountering the other particle again. The time needed for this is simply given by
\begin{equation}
T_{travel}=\frac{L}{2p\left(1-\frac{\omega_d}{\omega_a+\omega_d}\right)}\simeq\frac{L}{2p\left(1-\frac{\omega_d}{\omega_a}\right)}.
\end{equation}
The last factor of the denominator accounts for the time spent in the unbound state where no net displacement occurs as particles perform an unbiased random walk on the diffusive lane. The time for the exchange process $T_{exchange}$ can be calculated as follows: In a blocked configuration, one of the two particles will detach after a mean waiting time of $(2\omega_d)^{-1}$. The configuration is then as illustrated in figure~\ref{f:ptrap} and during the following sequence of moves, there is a trapping probability $p_{trap}$ (which we attempt to compute later on) that the unbound particle will reattach \emph{before} the bound particle was able to pass. The system is then again in the initial blocked configuration and the process has to start over by waiting on average another $(2\omega_d)^{-1}$. This leads to the following expression for the exchange time:
\begin{eqnarray}
T_{exchange}&=& \sum_{i=1}^\infty\frac{i}{2\omega_d}p_{trap}^{i-1}(1-p_{trap})\\
&=&\frac{1}{2\omega_d(1-p_{trap})^2},
\end{eqnarray}
where the time needed for the exchange itself has deliberately been neglected since the waiting time by assumption dominates all other time scales.

If we recall that a particle performs on average $L/2$ steps in between two blocked configurations, the average current per lattice site is
\begin{eqnarray}
\langle j^\pm\rangle&=&\frac{1}{L}\frac{L/2}{T_{travel}+T_{exchange}}\\
&=&\frac{1}{\frac{L}{p\left(1-\frac{\omega_d}{\omega_a}\right)}+\frac{1}{\omega_d(1-p_{trap})}}.
\label{eq:exchcurrent}
\end{eqnarray}
(Note that throughout the paper, angular brackets denote an average over stochastic histories.) This expression still depends on the trapping probability $p_{trap}$ for which we now derive an approximate value by rather intuitive considerations.

Let $P_u(X,t)$ (resp. $P_b(Y,t)$) be the time-dependent probability distributions to find the unbound (bound) particle $X$ ($Y$) lattice sites to the left of the starting configuration illustrated in figure~\ref{f:ptrap}. Then, $p_{trap}$ is the probability for the bound particle to take at most as many steps to the left as the unbound particle, summed over all times and distances travelled:
\begin{equation}
\label{eq:ptrap}
p_{trap}=\sum_{t=0}^\infty\sum_{x=0}^\infty\frac{1}{2}\omega_a\cdot P_b(Y\leq x,t)\cdot P_u(X=x,t).
\end{equation}
We have $x,t\in\NN$ as we consider discrete sites and time steps. The factor $\omega_a/2$ is necessary to assure that the unbound particle is indeed trapped by adsorbing to the filament.

The probability distribution for the unbound particle can be defined recursively by considering the possible moves:
\begin{eqnarray}
P_u(X=x,t=0)&=&\delta_{x0}\\
P_u(X=x,t>0)&=&D\cdot P_u(X=x-1,t-1)\nonumber\\
&&\quad+D\cdot P_u(X=x+1,t-1)\nonumber\\
&&\quad+(1-2D-\omega_a)\cdot P_u(X=x,t-1).
\end{eqnarray}
It is important to note that this distribution does not conserve the probability which reflects the increasing chance of the particle to reattach to the filament.

The particle in the bound state only has the choice to take a step forward or to stay on its site as we neglect the possibility of both particles to be simultaneously in the unbound state. The probability distribution is consequently a Poisson distribution with mean $pt$:
\begin{equation}
P_b(Y=x,t)=\frac{(pt)^x\exp(-pt)}{x!}.
\end{equation}

Using the above derived probability distributions, the value of \eref{eq:ptrap} can be computed numerically, yielding a value of
\begin{equation}
\label{eq:ptrapval}
p_{trap}=0.237341\ldots.
\end{equation}

\begin{figure}
 \begin{center}
 \includegraphics*[scale=0.5]{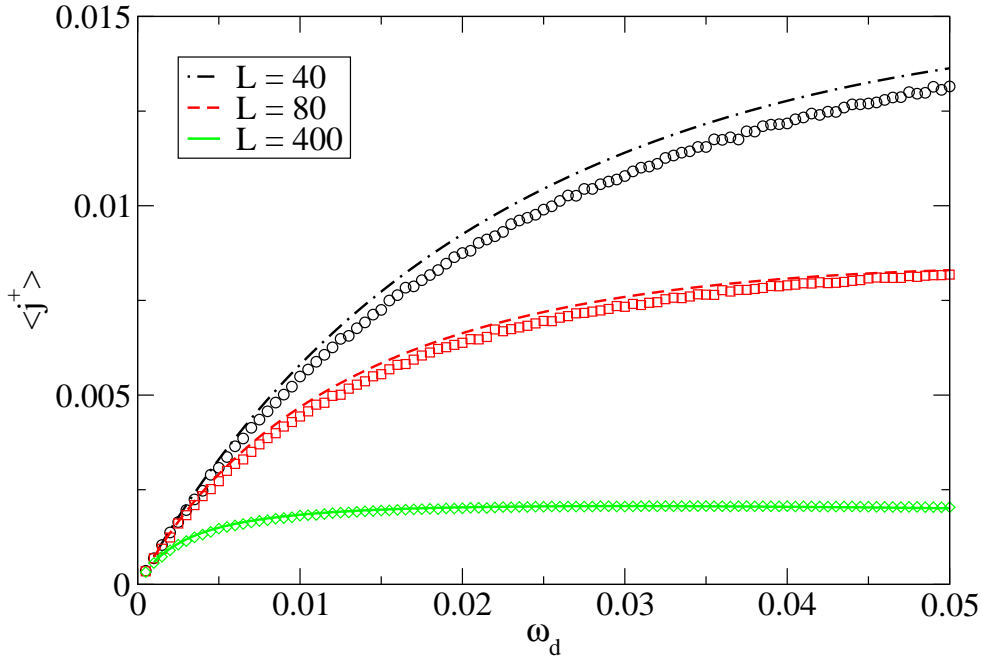}
 \caption{Average current $\langle j^+ \rangle$ along the microtubule filament in the two-particle system as a function of the detachment rate $\omega_d$ for different system sizes $L$ with the following set of parameters: $p=1$ and $\omega_a=D=0.33$. Dots are results from MC simulations, lines are the predicted behaviour by \eref{eq:exchcurrent}.}
 \label{f:2partexchange}
 \end{center}
 \end{figure}

Combining \eref{eq:ptrapval} with \eref{eq:exchcurrent}, we analytically derived an approximate expression for the current in a two-particle system at low detachment rates, which is well confirmed by numerical simulations (figure~\ref{f:2partexchange}). The current is systematically overestimated because the time needed for the exchange process has been neglected. At higher detachment rates, the assumption of never finding both particles in the unbound state does not hold any more so that the results lose their validity in this region of parameters. For small values of $\omega_d$, the travel times are dominated by the time needed for the exchange process. This parameter regime is relevant for processive molecular motors. Note that for many-particle systems with small but finite densities, it is not possible to consider solely two-particle clusters, since larger clusters form at any finite density. In this scenario, $T_{exchange}$ strongly depends on the cluster size and the arrangement of the particles, which complicates the analysis of the many particle system.

\section{Many-particle system}
\label{s:manypart}

In this section, we present an analysis of the many-particle case. The particle dynamics in the here considered interacting stochastic system is quite involved. Therefore we combine numerical simulations and a mean field approach in order to characterize the behaviour of the system.

Using the notations introduced in section~\ref{s:moddef}, the system of equations that has to be solved for a stationary state is given by
\begin{eqnarray}
\frac{\rmd\langle u_i^+\rangle}{\rmd t}&=&\omega_d\langle b_i^+\rangle+D\left(\langle u_{i+1}^+\rangle+\langle u_{i-1}^+\rangle\right)\nonumber\\
\label{eq:MEu}
&&\quad+\omega_a\langle u_i^+\left(1-b_i^+-b_i^-\right)\rangle-2D\langle u_i^+\rangle\\
\frac{\rmd\langle b_i^+\rangle}{\rmd t}&=&p\left[\langle b_{i-1,b}^+\left(1-b_i^+-b_i^-\right)-b_i^+\left(1-b_{i+1}^+-b_{i+1}^-\right)\rangle\right]\nonumber\\
\label{eq:MEb}
&&\quad+\omega_a\langle u_i^+\left(1-b_i^+-b_i^-\right)\rangle-\omega_d\langle b_i^+\rangle
\end{eqnarray}
with the corresponding equations for the negative particles. (In the following, we will restrict ourselves to write down the equations for the positive particles. The expressions for the negative particles are analogue.) On the right-hand side of equations \eref{eq:MEu} and \eref{eq:MEb}, we find the gain and loss terms of particles entering and leaving the considered local state $u_i$ or $b_i$ which make up the change of the occupations of these local states over time.

\subsection{Mean field approximation}
\label{ss:meanfield}

In order to find the physical properties of the stationary state, one would need to solve \eref{eq:MEu} and \eref{eq:MEb} with the temporal derivations set to zero ($\rmd/\rmd t=0$). Due to the complexity of the system, there is not much hope in finding an exact expression and we therefore have to resort to approximations.

By taking into account translational invariance of the system, an expression for a vertical equilibrium is found which expresses the equalilty of the number of particles adsorbing to and desorbing from the filament:
\begin{equation}
\omega_a\rho_u^+(1-\rho_b^+-\rho_b^-)=\omega_d\rho_b^+,
\end{equation}
where the replacements $\rho_u^+\equiv \langle u_i^+\rangle$, $\rho_b^+\equiv \langle b_i^+\rangle$ and the mean field approximation $\langle \tau\tau'\rangle=\langle\tau\rangle\langle\tau'\rangle$ ($\tau$ and $\tau'$ are arbitrary local states) have been applied. The densities $\rho_{u/b}^\pm$ do not depend on the lattice site since translational invariance is assumed and the mean field approximation thus provides a homogenous density profile. Additionally, we have an equation for the conserved total number of particles $\rho_u^++\rho_b^+=\rho_{tot}^+$ which in connection with the equations for the negative particles provides us with four equations for four variables. The total density in the system is consequently defined by the number of particles of a species divided by the system length $L$. The solution of this system of equations is the root of a quadratic expression:
\begin{eqnarray}
\label{eq:MF_Loesung}
\rho_{b}^\pm&=&\frac{\rho_{tot}^\pm}{2(\rho_{tot}^++\rho_{tot}^-)}\Bigg[\rho_{tot}^++\rho_{tot}^-+1+\frac{\omega_d}{\omega_a}\Bigg.\nonumber\\
&&\quad- \sqrt{(\rho_{tot}^++\rho_{tot}^-)^2+2(\rho_{tot}^++\rho_{tot}^-)\left(\frac{\omega_d}{\omega_a}-1\right)+\left(\frac{\omega_d}{\omega_a}+1\right)^2}\Bigg].
\end{eqnarray}
With this solution, we find the stationary current in the system to be
\begin{equation}
\langle j^+\rangle=p\rho_b^+(1-\rho_b^+-\rho_b^-).
\end{equation}

Since the mean field approximations neglects all correlations between individual particles, we do not expect it to yield good results when blocking situations with particles of different species are frequent. This will be the case if the ratio of desorption to adsorption rate $r\equiv\omega_d/\omega_a$ is small, since particles will prefer the bound state. Furthermore, the mean field approximations will give poor results, if we have very low desorption \emph{and} adsorption rates independent of their ratio. In this case, particles spend a lot of time on the same lane so that correlations have enough time to build up. Therefore the validity of the mean field approach is restricted to parameter regimes where the particle dynamics is dominated by diffusion.

\begin{figure}
 \begin{center}
 \includegraphics*[scale=0.5]{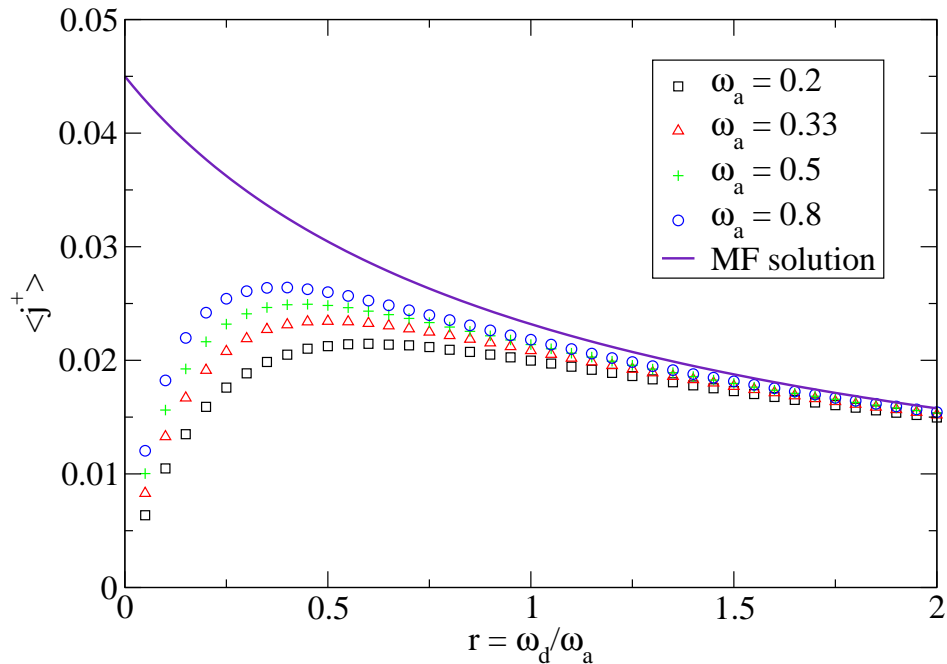}
 \caption{Average current $\langle j^+ \rangle$ along the microtubule filament in a system with many particles as a function of the ratio of detachment to attachment rate $r=\frac{\omega_d}{\omega_a}$. The continued line is the mean field solution and the dots are the results from MC simulations. The set of parameters used is $L=1000$, $\rho^\pm_{tot}=0.05$, $p=1$, and $D=0.33$.}
 \label{f:MFcurrent}
 \end{center}
 \end{figure}

These predictions are well confirmed by the results from MC simulations, as can be seen in figure~\ref{f:MFcurrent}. The important difference of the results when compared to the mean field predictions for small $r$ come from the jam-like accumulations of particles on the filament at low detachment rates. Remembering the biological motivation of the model, it is exactly this regime of low values for $\omega_d$ that is of interest. The next subsection will consequently be devoted to the characterization of these jams that will be called clusters in the following.

\subsection{Clustering}
\label{ss:clustering}

\begin{figure}
 \begin{center} 
 \includegraphics*[scale=0.5]{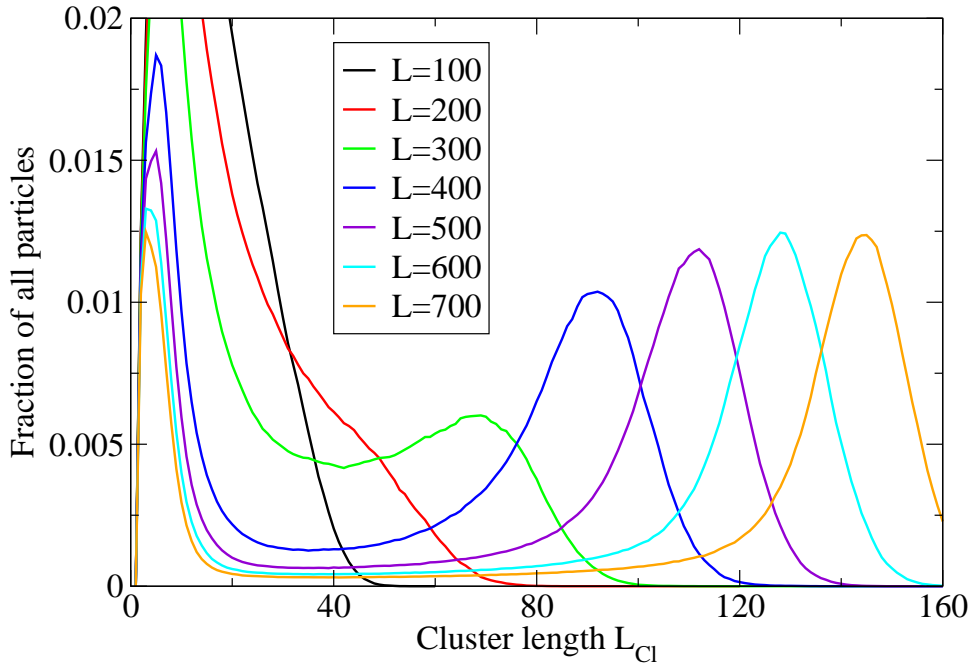}
 \caption{Distribution of cluster appearance for the standard set of parameters in a system of density $\rho_{tot}^\pm=0.3$. Large clusters appear if the system is large enough and enough particles are in the system ($L\gtrsim 300$).}
 \label{f:clusterdistr}
 \end{center}
 \end{figure}

For the following investigation of clustering within this model, we make use of cluster size distributions such as shown in figure~\ref{f:clusterdistr}, where the fraction of all particles in a cluster of a certain length is drawn as a function of the cluster size. For our analysis, a cluster is defined as an accumulation of particles on the filament with only single empty filament sites within the cluster. We checked carefully that alternative cluster definitions do not alter the results qualitatively. If not stated otherwise, the standard set of parameters used in the following was $\omega_d=0.02$, $\omega_a=D=0.33$, and $p=1$.

A first result is that for high enough particle \emph{numbers}, there is a transition from a well-mixed phase with only very short clusters to a phase in which a single large cluster builds up and dominates the behaviour of the whole system (see figure~\ref{f:clusterdistr}). As the mean of the cluster size distribution shifts towards shorter clusters and the variance increases for shorter system sizes $L$, the actual onset of the clustering is hard to determine because the fluctuations in the cluster length are of the same magnitude as the cluster length itself.

\begin{figure}
 \begin{center} 
 \includegraphics*[scale=0.5]{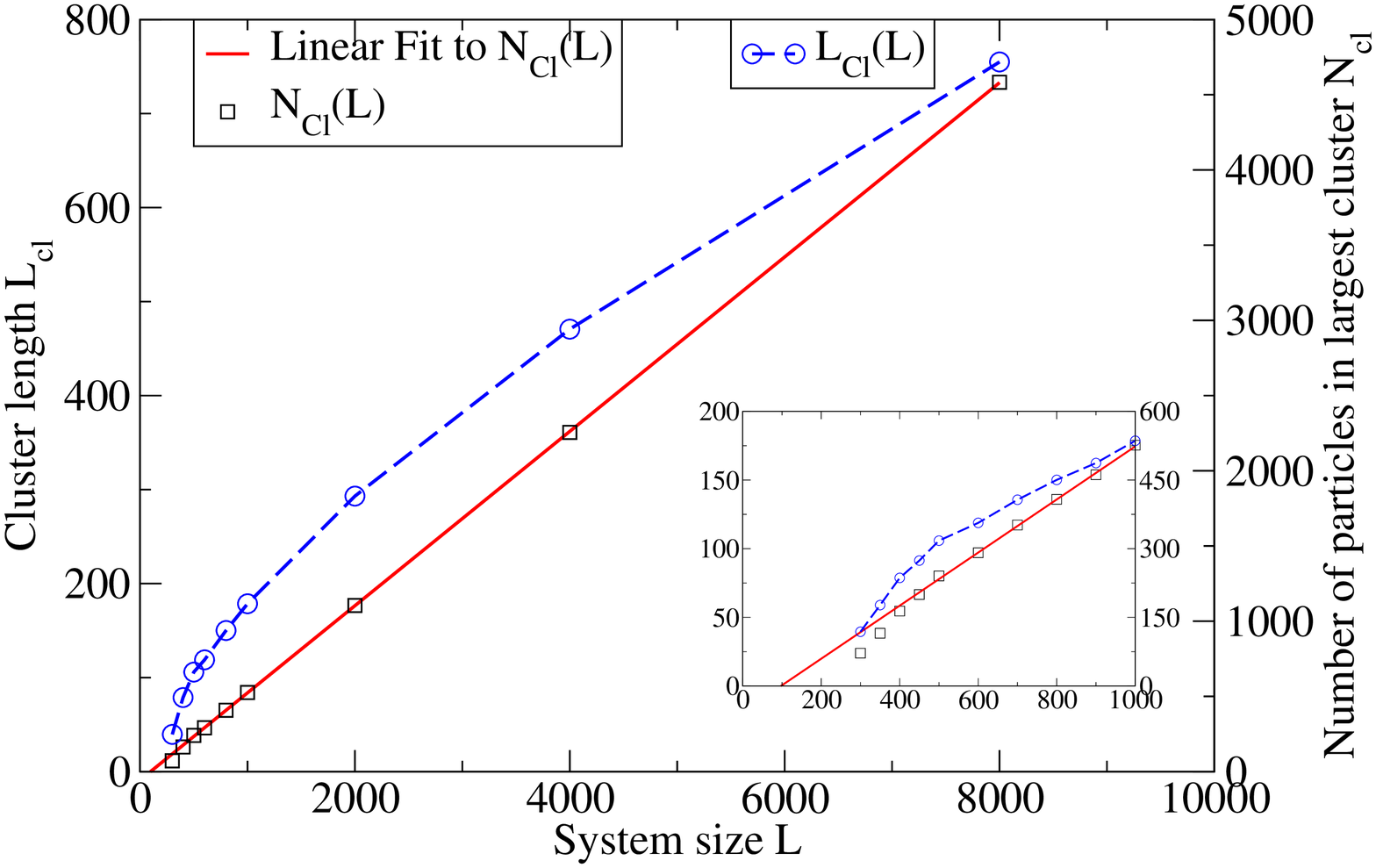}
 \caption{Total number of particles involved in the largest cluster $N_{Cl}$ (black squares) and mean cluster length $L_{Cl}$ (blue circles) as a function of the system size $L$ for $\rho_{tot}^\pm=0.3$. $N_{Cl}$ grows linearly for large $L$ whereas the cluster length shows sublinear growth. The linear fit equation (red line) is $N_{Cl}=0.57954L-56.188$.}
 \label{f:Ncl}
 \end{center}
 \end{figure}

On the other hand, the fluctuations become negligible for large systems, which enables us to make observations valid for the thermodynamical limit. When increasing the system size $L$ while keeping the particle density $\rho_{tot}^\pm$ constant, the peak in the cluster distribution $L_{Cl}$ shifts sublinearly to greater cluster lengths (see figure~\ref{f:Ncl}) and decreases in size (graph not shown). The fact of the decreasing impact of the cluster raises the question where the particles go, because a decreasing fraction of particles in the cluster could mean that the clustering disappears for very large systems.

\begin{figure}
 \begin{center} 
 \includegraphics*[scale=0.5]{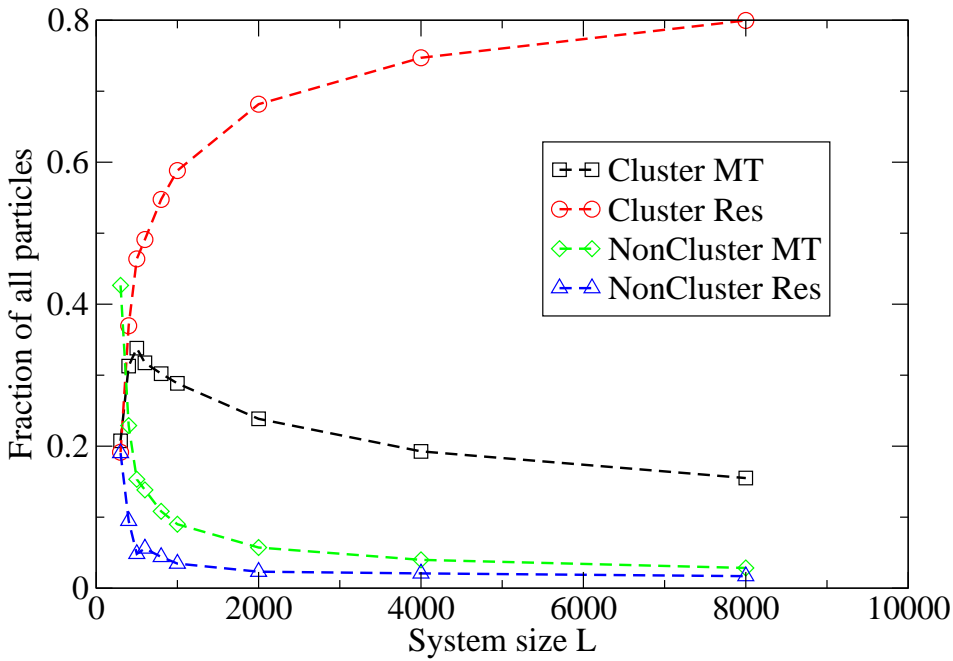}
 \caption{Fraction of particles in the four different regions of the system for different system sizes $L$ and constant density $\rho_{tot}^\pm=0.3$. Red circles stand for particles in the diffusive lane above the large cluster, black squares for particles in the cluster, green diamonds for particles on the filament but outside the largest cluster, and blue triangles for particles in the diffusive lane that are not above the large cluster.}
 \label{f:partfrac}
 \end{center}
 \end{figure}

In order to draw the curves of figure~\ref{f:partfrac}, a clustering system with density $\rho_{tot}^\pm=0.3$ has been subdivided into four regions: the largest cluster on the filament as defined above, the sites in the diffusive lane next to the cluster, the filament sites that do not belong to the largest cluster, and the corresponding  sites on the diffusive lane. The black squares in figure~\ref{f:partfrac} thus correspond to the area beneath the large peak in the cluster size distribution. We observe that the involved particles accumulate in the diffusive lane above the large cluster. As there is no interaction, sites are multiply occupied. A further investigation of the cluster properties yields an almost perfect linear relation between the total number of particles involved in a cluster $N_{cl}$, i.e., in both lanes, and the system size $L$: $N_{cl}=A\cdot L-B$. Figure~\ref{f:Ncl} shows that this relation is consistent with the numerical data. Deviations are only observed for small $L$ where fluctuations destabilize the largest cluster (see, e.g., the cluster size distribution for $L=100$ in figure~\ref{f:clusterdistr}). On the other hand, the above equation turns out to be validated very well in the limit $L\to\infty$ where we get \eref{eq:clusterfraction} for the fraction of particles in the cluster,

\begin{equation}
\label{eq:clusterfraction}
\frac{N_{cl}}{L\cdot(\rho_{tot}^++\rho_{tot}^-)}=\frac{A}{\rho_{tot}^++\rho_{tot}^-}-\frac{B}{\rho_{tot}^++\rho_{tot}^-}L^{-1},
\end{equation}
thus deriving that in the thermodynamical limit the fraction $A/(\rho_{tot}^++\rho_{tot}^-)$ of all particles will condensate in the largest cluster. This number turns out to be near but still smaller than 1. Consequently, the cluster takes up a finite fraction of the particles in the limit of large system sizes. The offset in the linear equation describes well the scaling behavior when compared to numerical data.

The fraction of particles in the cluster approaches an asymptotic value for large system sizes, but the cluster length grows sublinearly with the number of particles, which is due to the increase in the occupation of the reservoir sites. The almost constant fraction of particles in the large cluster means that the same fraction of particles that are not involved in the cluster has to be distributed over an increasing part of the system. This leads to a decreasing density outside the cluster. Furthermore, the current and the density are positively correlated for the low densities that are found in the homogenous regions outside the large cluster. This indicates a decreasing current in the presence of larger clusters. But in this case, the current in the system can be seen as the outflow of the cluster ($j_{Cl}=\langle j^+\rangle+\langle j^-\rangle$), as the cluster represents a big obstacle for any particle. So we can establish a relation between the outflow of a cluster and its number of particles.

\begin{figure}
 \begin{center} 
 \includegraphics*[scale=0.5]{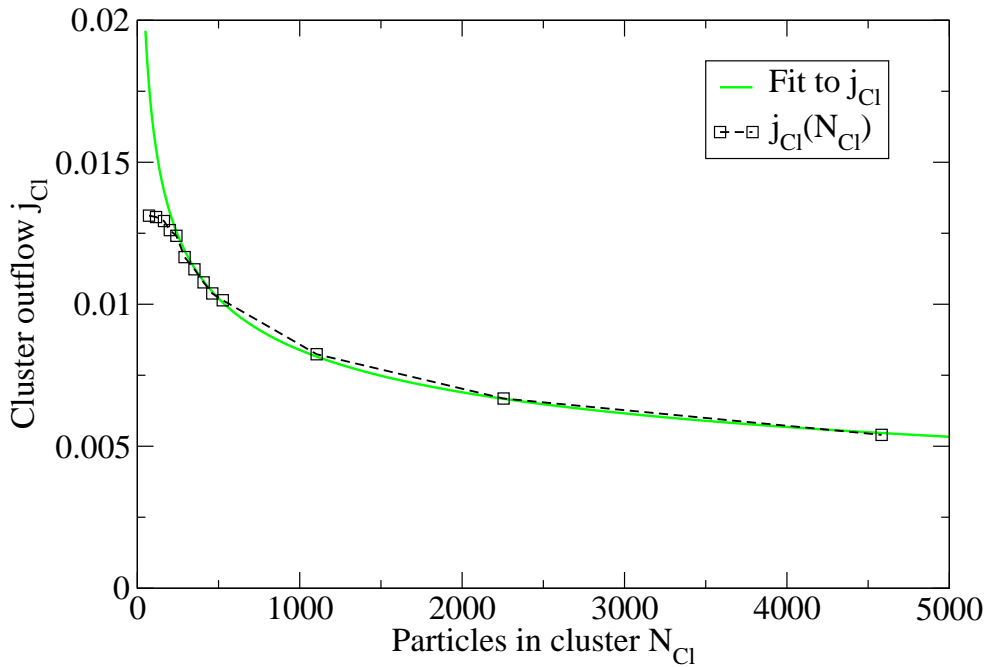}
 \caption{Total current in the system as a function of the number of particles in the largest cluster for $\rho_{tot}^\pm=0.3$. The green line is a fit of the function $j_{Cl}(N_{Cl})=j_\infty\left(1+\frac{b}{N_{Cl}^\sigma}\right)$ that gives the parameters $J_\infty=0.00012$, $b=511$, $\sigma=0.29$.}
 \label{f:clusteroutflow}
 \end{center}
 \end{figure}

Analysing the outflow from the large cluster enables us to check the criterion for phase separation in a one-dimensional system introduced by \emph{Kafri et al.}~\cite{Kafri02}. It relies on the asymptotic behaviour of the current out of a domain of a certain size. For the application of this criterion to the treated model, the domain size is here identified as being the number of particles in the cluster $N_{Cl}$. Consequently, we try a fit of the function
 \begin{equation}
j_{Cl}(N_{Cl})=j_\infty\left(1+\frac{b}{N_{Cl}^\sigma}\right)
\end{equation}
to our data for the particle current in the system as shown in figure~\ref{f:clusteroutflow}. Here, we are interested in the exponent $\sigma$ which determines whether phase separation occurs. In the case of $\rho_{tot}^\pm=0.3$ its value is $\sigma=0.286\pm0.062$, thus clearly below 1, resulting in a phase separation at any density in the thermodynamical limit~\cite{Kafri02}. Again, the data for small system sizes deviates from the assumed behaviour which does not have any influence on the result for large scales.

The influence of the other parameters has also been investigated. We will not carry out the analysis in this paper, but the results can be briefly summarized as follows:

Decreasing the desorption rate $\omega_d$ or increasing the adsorption rate $\omega_a$ leads to larger clusters containing an increasing fraction of particles. If desorption is too strong or adsorption too weak, cluster formation is inhibited and the system is no longer in the region of low ratio $r=\omega_d/\omega_a$ which means that the mean field solution regains validity.

A higher rate of diffusing moves $D$ leads to a less sharply peaked cluster distribution and shifts the maximum to higher cluster lengths. Both effects can easily be understood as the diffusion controls the outflow of a cluster and stronger diffusion will consequently disperse the particles in the diffusive lane over more sites. Note that coupling to a bulk reservoir as in~\cite{Parmeggiani03} corresponds to the limit $D\to\infty$. In this case, no large clusters will appear since particles lose all memory about the site at which they left the bound state.

When keeping the system length $L$ fixed and varying the number of particles in the system by increasing $\rho_{tot}^\pm$, the results are very similar to figure~\ref{f:partfrac}.

An important result of this investigation is the high robustness of the clustering phenomenon to parameter variation.

\subsection{Biologically relevant parameters}
\label{ss:bio}

We briefly discuss the behaviour of the model for parameter values that are in the biologically relevant scale. The choice of the numerical values followed the data chosen in \cite{Mueller08}. Because of the need of symmetric parameters (i.e., rates that do not depend on the particle species), only the orders of magnitude of the rates have been kept and then rescaled in order to have $p=1$. The rate $D$ has been calculated by considering a one-dimensional random walk along the diffusive lane which gives us a connection to the real physical diffusion constant $K_D$ ($K_D=\frac{\Delta x^2}{2\Delta t}$). The diffusion constant for a spherical vesicle of about $100~nm$ in radius is given by
\begin{equation}
K_D=\frac{k_BT}{f} \qquad \mathrm{with} \qquad f=6\pi\eta r.
\end{equation}
The cytoplasm's viscosity is given to be $\eta=3000~mPa/s$ \cite{Mogilner96}. Note that the diffusional processes are dominant when considering single motor proteins without attached cargo. In this scenario, $D$ would take larger values than the stepping rate $p$. The attachment rate clearly depends on the geometry of the considered system. In~\cite{Mueller08}, the rate is given under the condition that the motor protein is already located near the filament. This is not necessarily the case for our model where we see $\omega_a$ rather as an effective attachment rate to account for possible diffusion in radial direction. Consequently, the given value represents an upper bound for a single motor which we will assume here because of the relatively dense packing in an axon with its several microtubules in parallel. In fact, a value of about $100~Fil/\mu m^2$ has been observed in rat embryos~\cite{Thies07}. Altogether, we obtain the set of parameters in table \ref{t:biopara}.

\begin{table}[h]
\caption{Table of approximate values for biologically relevant parameters. Orders of magnitude have been chosen as in~\cite{Mueller08}. References are given in the table. The rescaled time unit is the average time for a forward step of $8~nm$ in order to obtain $p=1$.}
\label{t:biopara}
\begin{indented}
\item[]\begin{tabular}{@{}lrr}
\br
Parameter&Approximate value&Rescaled value [$(0.01~s)^{-1}$]\\
\mr
Stepping rate $p$~\cite{Vale96,Carter05,King00,Nishiura04}&$0.8~\mu m/s$&$1.0$\\
Desorption rate $\omega_d$~\cite{Vale96,Schnitzer00,King00,Mallik04}&$1~s^{-1}$&$0.01$\\
Adsorption rate $\omega_a$~\cite{Leduc04,King00,ReckPeterson06}&$5~s^{-1}$&$0.05$\\
Diffusion rate $D$&$10.0~s^{-1}$&$0.1$\\
\br
\end{tabular}
\end{indented}
\end{table}

In a system of length $L=1000$, we find the same clustering effects as before for rather low densities ($\rho_{tot}^\pm\geq0.2$).

\section{Coupling of two systems}
\label{s:coupledsys}

A microtubule bundle offers the motor proteins more than one filament to which they can bind. In this paper, this has not been taken into account so far and might be a crucial improvement of the model as the possibility of changing from one protofilament to another might prevent cluster formation. These sideward steps have been observed experimentally at least for dynein~\cite{Wang99,Ross06}.

To include this, we extend the model to be constituted of two subsystems defined as in section~\ref{s:moddef}. The coupling of the subsystems is assured by the possibility of a filament change with rate $c_{MT}$ if the next site in the stepping direction is occupied. The two diffusive lanes representing the cytoplasm are coupled by a reservoir change rate $c_R$ which is not subject to any other condition. The additional moves are formally stated in table \ref{t:movescoupled}.
\begin{table*}
\caption{Table of additional moves in the coupled system ($n,m\in\NN,\ n\geq 1,\ j\neq k$; $j$, $k$ denoting the subsystem)}
\label{t:movescoupled}
\begin{indented}
\item[]\begin{tabular}{@{}lll}
\br
Move & Rate & Biological interpretation \\
\mr
$\{b_{i,j}^+=1,b_{i+1,j}^++b_{i+1,j}^-=1,b_{i,k}^+=0,b_{i,k}^-=0 \}$& $c_{MT}$ & Filament change of\\
\quad$\to \{b_{i,j}^+=0,b_{i+1,j}^++b_{i+1,j}^-=1,b_{i,k}^+=1,b_{i,k}^-=0 \}$ &&\quad kinesin when blocked \\
\mr
$\{b_{i,j}^-=1,b_{i-1,j}^++b_{i-1,j}^-=1,b_{i,k}^+=0,b_{i,k}^-=0 \}$& $c_{MT}$ & Filament change of \\
\quad$\to \{b_{i,j}^-=0,b_{i-1,j}^++b_{i-1,j}^-=1,b_{i,k}^+=0,b_{i,k}^-=1 \}$ &&\quad dynein when blocked \\
\mr
$\{u_{i,j}^\pm=n,u_{i,k}^\pm=m \}$&$c_{R}$ & Lateral diffusion \\
\quad$\to \{u_{i,j}^\pm=n-1,u_{i,k}^\pm=m+1 \}$ &&\quad around the MT \\
\br
\end{tabular}
\end{indented}
\end{table*}
Rates are still taken to be symmetric for both particle species. This symmetrization does not correspond to the biological situation as there is no experimental evidence that kinesins are able to jump from one protofilament to another. However, we made sure that allowing only one particle species in our model to change from one subsystem to another has very little effect on the clustering properties of the system and simply leads to species-dependent currents, causing a net current in the system.\\

\begin{figure}
 \begin{center} 
 \includegraphics*[scale=0.5]{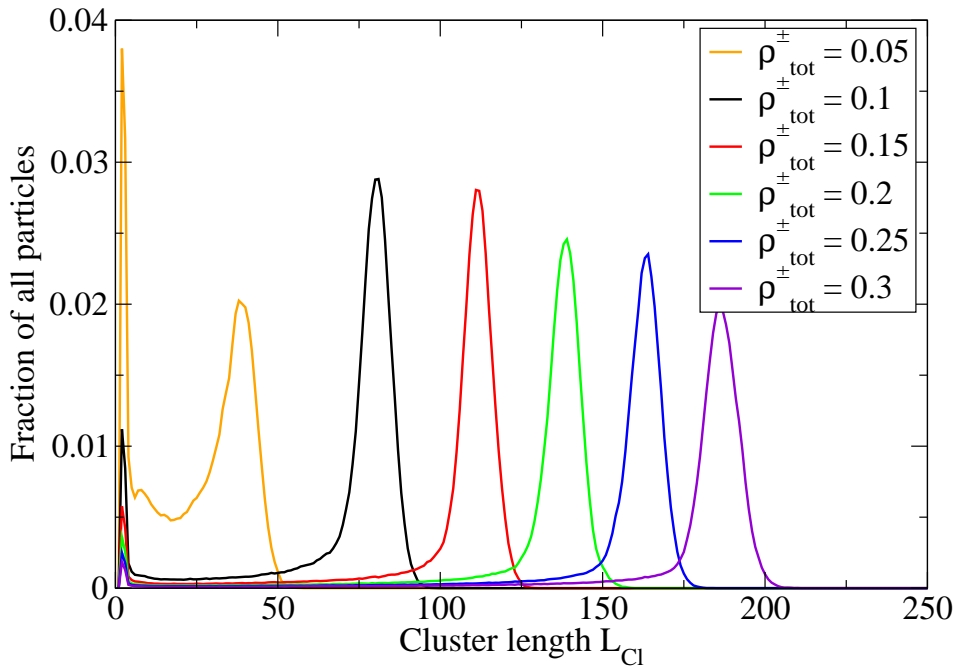}
 \caption{Distribution of cluster appearance for the standard set of parameters in a coupled system of length $L=1000$. The coupling is assured only by filament changes: $c_{MT}=0.1$, $c_R=0$.}
 \label{f:clusterdenscoupled}
 \end{center}
 \end{figure}

At first sight counter-intuitively, cluster formation is strongly promoted by the coupling of the two subsystems and will appear even at very low global particle densities as can be seen in figure~\ref{f:clusterdenscoupled}. Although the data shown here has been produced by only allowing inter-filament moves while $c_R=0$, we verified that an additional coupling of the reservoirs by a rate $c_R\neq0$ does not significantly change the cluster distribution. As a general result, we obtained that the promotion of clustering does not depend on the way of coupling. Yet the inter-filament changes are more efficient, which means that an uncoupled system without large clusters starts clustering at lower filament change rates $c_{MT}$ than reservoir change rates $c_R$ needed to induce clustering.

The coupling causes an accumulation of particles in a subsystem to be transported to the other subsystem rather than to the reservoir where more cluster outflow and thus a destabilizing factor would be generated. The higher concentration of particles in the second subsystem then most likely leads to a cluster parallel to the first one. Once this situation is obtained, the clusters stabilize themselves by reciprocally filling holes potentially left by detaching particles and adjusting their lengths. Thus, the peaks in the distribution of cluster sizes are much sharper than in the uncoupled system.

An investigation of the coupled model with biological parameters is difficult due to the uncertainty of the filament change rate $c_{MT}$ for which we were unable to find any numerical value. By contrast, the reservoir change rate can be estimated in a similar way as has been done for the diffusion rate $D$ by using the lateral distance of two neighboring protofilaments that can be approximated by using the fact that 13 protofilaments are arranged to form a cylinder of a diameter of $25~nm$~\cite{Alberts02}. In any case, the above result of clustering will persist and clusters will appear at even lower densities than without coupling.

A generalization to more filaments does not modify the above observation but rather leads to even sharper maxima in the distribution of cluster sizes for the same reasons as explained above.

\section{Discussion}
\label{s:discussion}

We introduced a bidirectional stochastic lattice gas model based on the microtubular traffic within axons driven by motor proteins. We considered two types of particles moving in opposite direction on the filament. By using this largely simplified description of intracellular traffic along microtubules we tried to gain a better insight into the elementary processes determining the transport capacity of the system. 

The model we use in this work can be interpreted as a modification of the model in \cite{Klumpp03}; the major modification being the introduction of a second particle species identical to the first one except for its preferential moving direction when bound to the filament. The geometry of the model remains unchanged, i.e., the axon is reduced to a one-dimensional lattice with two lanes, one representing the filament and the other one representing the surrounding cytoplasm. The only possibility to bypass obstacles is to pass through the interaction-free cytoplasm.

Models in which particles of different polarity switch sites with a diminished rate exist, e.g. \cite{Arndt98,Korniss99}. In consequence, these models do not need a cytoplasmic lattice. These types of models summarize the complex process of detachment, diffusion and reattachment behind the obstacle in a constant rate of site exchange. In the model presented in this paper, the behaviour differs qualitatively due to the fact that motion in the diffusive lane is not directed and particles do not cross an existing cluster with a low but constant velocity. This leads to a cluster outflow that heavily depends on the length of the cluster. This implies that current-density relations, i.e., fundamental diagrams, depend strongly on the system size, in contrast to the findings of~\cite{Arndt98,Korniss99}.

As the system's behaviour is dominated by situations in which particles of different species block each other mutually, we first treated analytically the process of two particles exchanging their sites on the lattice and came to satisfying results in the regime of low detachment rates $\omega_d$. The calculations were carried out by deriving approximate time-dependent probability distributions of the particle positions and the solution well reflected the numerically observed increase in current with increasing detachment rate.

In the case of more than one particle of each species, a mean field analysis was only possible in the limit of high desorption rates. Then, particles demonstrate a high affinity for the diffusive lane where correlations cannot build up. In this regime, the system behaves somewhat fluid-like with few jamming situations and it is mostly conditioned by the equilibrium between filament and reservoir occupation.

If the particles' affinity to the filament is high, jamming occurs. MC simulations have shown that a transition to a regime with a single big cluster exists for large enough particle numbers and low enough detachment rates. The similar behaviour when adding particles either by keeping the global density constant and increasing the system size or by leaving the system size constant and increasing the global density suggests a dependence of the clustering effects on the number of involved particles and not on particle density. Effectively, analyzing either the total fraction of particles in the large cluster or the outflow of the largest cluster, numerical evidence is gained that back up the existence of cluster formation at all densities in the thermodynamic limit.

The coupling of the filaments with a very short-ranging (next-neigbour) interaction does not lead to formation of tracks mainly occupied by a single particle species. Instead, the coupling even enhances clustering in the system.

All in all, the model shows a strong tendency towards clustering. The effect of unbound particles ``remembering'' their previous position in the bound state is substantial for this cluster formation. This lets us conclude that the presented results are generic for systems with confined geometries that induce this kind of memory. The related model by Parmeggiani et al.~\cite{Parmeggiani03} is obtained for $D\to\infty$ in which case clustering vanishes. We consider our assumption of small $D$ to be more relevant for the biological situation, especially in the context of axonal transport.

The accumulation of axonal cargo is obviously not a physiologically desired phenomenon. Considering the length of axons (up to $1m$), the particle density needed in order to have enough particles to form a stable cluster are very low, thus causing a constant risk of clustering. This effect will be even stronger if exclusion effects on the diffusive lane are taken into account, which have been neglected here. This leads to the conclusion that another mechanism has to be incorporated if one aims at modeling intracellular transport, because accumulations of axonal cargo are not observed in healthy neurons. The transport is in fact very efficient and oppositely moving vesicles or organelles are not seen to hinder each other. This would be a strong argument for track formation within the biological system.

Furthermore, the assumption of periodic boundary conditions clearly does not reflect the biological situation. In our work, we were interested in the formation of clusters, which is not a boundary effect and was shown to represent the generic behaviour of our model. Introducing open boundary conditions might lead to a subtle interplay between the cluster dynamics and the boundaries, which will be investigated in future work.

\vspace{\baselineskip}

\ack

We thank R J Harris for valuable discussions and the DFG Research Training Group GRK 1276 for financial support.

\section*{References}
 \bibliographystyle{unsrt}
 
 \bibliography{references.bib}

\end{document}